# Non-Ionizing Radiation Analysis in Close Proximity to Antenna Tower: A Case Study in Northeast Brazil

Andre B. de F. Diniz[1], Vicente A. de Sousa Jr.[1], Marcio E. C. Rodrigues[1], Halysson B. Mendonça[2], Gutembergue Soares da Silva[1], Fred Sizenando Rossiter Pinheiro[1]
[1]*Federal University of Rio Grande do Norte (UFRN), Natal, Brasil, RN – andrebfd@ufrn.edu.br, vicente.sousa@ufrn.edu.br, mecrodrigues@gmail.com, guttembergue@gmail.com, fredrossiter@uol.com.br,*
[2]*Agência Nacional de Telecomunicações, Natal, RN, Brazil - halysson@anatel.gov.br*

***Abstract*—While the amount of telecommunications services grows rapidly in the whole world, humans get potentially more exposed to Non-Ionizing Radiation (NIR) from a number of different sources. Measurements of NIR levels are relevant in order to compare the results with national and international standards, aiming at the preservation of human health. Thus, it is of great interest to explore a variety of topics regarding this subject. Based on this need, this paper has a number of goals, including monitoring radiation levels in an everyday situation, an investigation of how national and international regulations organs address NIR levels and a demonstration of scientific production trends regarding this topic. The work also presents a bibliometric study about the main scientific productions and trends related to NIR measurements, with focus on the field of Telecommunications. The analysis is based on the databases of Web of Science (WOS) and Scopus, computing publications in the last 10 and 3 years, showing a trend evaluation. Among the main results from this exploratory investigation, there is an exponential growth in the number of publications about NIR, ranging from research in physics to medicine. The way research contribution in different countries is also shown, in order to support the relevance of the topic in defining new tendencies for new investigations. Furthermore, the NIR exposure limits and measurement criteria are presented both in Brazilian and in international norms, demonstrating how different national regulations can be when compared to international guidelines. Another contribution is a case study in Natal, Brazil, where NIR levels were monitored in four distinct locations over a period of 24 hours each and compared to the current regulations. The result is a quantitative analysis of the amount of radiation that some populations might be exposed to in distinct moments of the day. Measurements were carried out in the proximity of an antenna spot, since it could represent a source of high NIR levels (a worst case of human exposure). This campaign differs from older studies by including a more intense usage of microwaves frequencies due to operating 4G and pre-5G systems in Brazil. Finally, a statistical study involving the measurements is conducted, concluding the analysis of how some population groups might be affected by NIR.***

# I. INTRODUCTION

Non-Ionizing Radiation (NIR) designates electromagnetic fields (EMF) having enough energy to only raise the state of electron excitation, without ionizing atoms [1]. Its sources can be both natural (solar radiation, electric discharges and others) and artificial (antennas, power lines and others), and its effects are highly dependent on the frequency of the waves.

There is a constant international concern about the consequences of exposure to NIR, since, according to the literature, there is the possibility of triggering harmful effects to human health [2]. Based on this, several organizations have established guidelines for exposure limits, including, for example, ICNIRP (International Commission on Non-Ionizing Radiation Protection) in a global context [1], and ANATEL (Brazilian National Telecommunications Agency) on the Brazilian national scene [3], [4]. Eventually, such agencies update their rules as the spectrum usage evolves and as increasing knowledge about potential human health effects, due to NIR exposure, causes revisions on exposure level limits.

Given the accelerating rise of communication services and the prospects of an even more connected society, it is of crucial importance to monitor the radiation levels which the general population might be exposed to, comparing them to the current regulations, especially when exposure limits or measurement procedures have changed in some way. Additionally, the analysis on how scientific publications are distributed in the world is important, in order to map and report the tendencies regarding NIR studies.

A crucial aspect regarding NIR investigation is the target bandwidth for measurements. Table I shows the Radio Frequency (RF) bands defined by the International Telecommunications Union (ITU) [5]. Nowadays, more attention is being given to frequencies ranging from 300 MHz - lower end of Ultra High Frequency (UHF) - to 30 GHz - upper end of Super High Frequency (SHF) -, since that is where most of the telecommunication services operate, including the incoming 5G networks.

TABLE I. NOMENCLATURE FOR DIFFERENT FREQUENCY RANGES [5].

| Frequency range | Nomenclature |
| --- | --- |
| 300Hz to 3kHz | Ultra Low Frequency (ULF) |
| 3kHz to 30kHz | Very Low Frequency (VLF) |
| 30kHz to 300kHz | Low Frequency (LF) |
| 300kHz to 3MHz | Medium Frequency (MF) |
| 3MHz to 30MHz | High Frequency (HF) |
| 30MHz to 300MHz | Very High Frequency (VHF) |
| 300MHz to 3GHz | Ultra High Frequency (UHF) |
| 3GHz to 30GHz | Super High Frequency (SHF) |
| 30GHz to 300GHz | Extremely High Frequency (EHF) |

## A. Related Works

A number of publications provided by ICNIRP contributed to the definition of the recommended maximum exposure levels [1]. They consider a wide variety of aspects regarding biological phenomena in relation to electromagnetic waves, such as energy absorption and coupling to low-frequency electric and magnetic fields [6], [7].

There is a considerable amount of literature addressing effects of NIR exposure and how the human body reacts to it in different situations [8]–[10]. Although most of the achieved results are inconclusive, it is still of interest to monitor these levels, given the rise of sources which the population may be exposed to in the future.

Diversified exposure scenarios exist for different frequency ranges and size of sources. Thus, measurements must be held in particular manners for each case; some of them are explored in larger (power lines) [11], medium (indoor power substations) [12] and smaller (smart glasses) [13] scale, also pointing to the necessary apparatus. Evaluation of NIR exposure at street level can be found in the literature and some contributions have been made recently in a number of contexts, including: FM/TV broadcasting services [14] and outdoor urban environments in different cities [15]–[18].

However, there is a lack of contributions at skyline level of apartment buildings close to antenna towers. Considering the radiation pattern of typical urban telecommunication services, we claim this is a worst-case situation, but very common in verticalized cities. Additionally, some relevant elements, to the best of the authors' knowledge, are not found in previous works, such as surveys on NIR publications and discussions on NIR measurement procedures. In view of the given situation, the key contributions of this paper are:

- Analysis of scientific production about NIR in the Web of Science (WOS) and Scopus databases, looking for trends and discussing the investment regions, publications by years, and by area of knowledge;
- Brief explanation of relevant topics on the Brazilian regulation about NIR exposure assessment and comparison of this regulation with the ICNIRP guidelines, regarding NIR exposure levels, highlighting the differences between them and how they have changed;
- Presentation and discussion of a case study results in Natal, Brazil, with measurements at skyline level of apartment buildings close to antenna towers, where several NIR measurement rounds were conducted over 24 hours in each spot;
- Evaluation of the outcomes from the measurements under national and international regulations, filling a current need in the literature due to the massive adoption of 4G services in the target region for measurements as well as the recent changes in the Brazilian regulations regarding NIR exposure levels, also providing a means for further comparisons.

The paper is presented as follows. Section II exposes the methodology for the bibliometric research and its results, also analyzing its outcomes. Section III shows international and Brazilian regulations regarding NIR exposure limits, alongside a discussion on their updates and the differences between them. In Section IV, a case study is presented in terms of setup, methodology, field intensity estimation, evaluation, and discussion of the achieved results. Finally, Section V contains conclusions and final remarks about the topics discussed in the paper, as well as some further study suggestions.

## II. Bibliometric Research

Given the previously described problem, the present study seeks firstly to analyze scientific production on NIR and the perspective of continuity for the next years, looking for trends. To accomplish this goal, a bibliometric research was carried out.

### *A. Background Research Methodology*

The Web of Science (WOS) and Scopus databases were used to carry out the prospective study on trends in NIR. The research was based on the title of the publications, from the first found publication about NIR until 2008 and in the period of the last 10 years (2009-2019). The keywords used were "Non-ionizing" and "Radiation". The search focus only on the title is justified by the vast amount of publications not directly related to the effects of NIR and its measurement, extending from fields such as physics and medicine, which can be verified through the search for the topic (title, abstract and keywords).

To analyze the production in terms of area of knowledge (Table II), the 50 most cited articles were considered, with the classification of areas being made by reading the title and abstract of the publications. In the search for keywords, those that did not represent a technique, technology or an application area were excluded.

All graphics presented in this section are based on numbers collected at the end of January of 2020, thus, including the whole data of 2019 publications.

### *B. Time analysis*

The first publications registered in the databases date from 1945 (WOS) and 1940 (Scopus). Since then up to the present year, there have been more than 2,642 publications via WOS and 2,551 publications via Scopus.

One of the first analyzes in a bibliometric study is that of the development of research over the years. Fig. 1 shows the number of publications per year, from 1946 to 2016, considering 10-year intervals; from that point on, the number of publications in 2019 was analyzed.

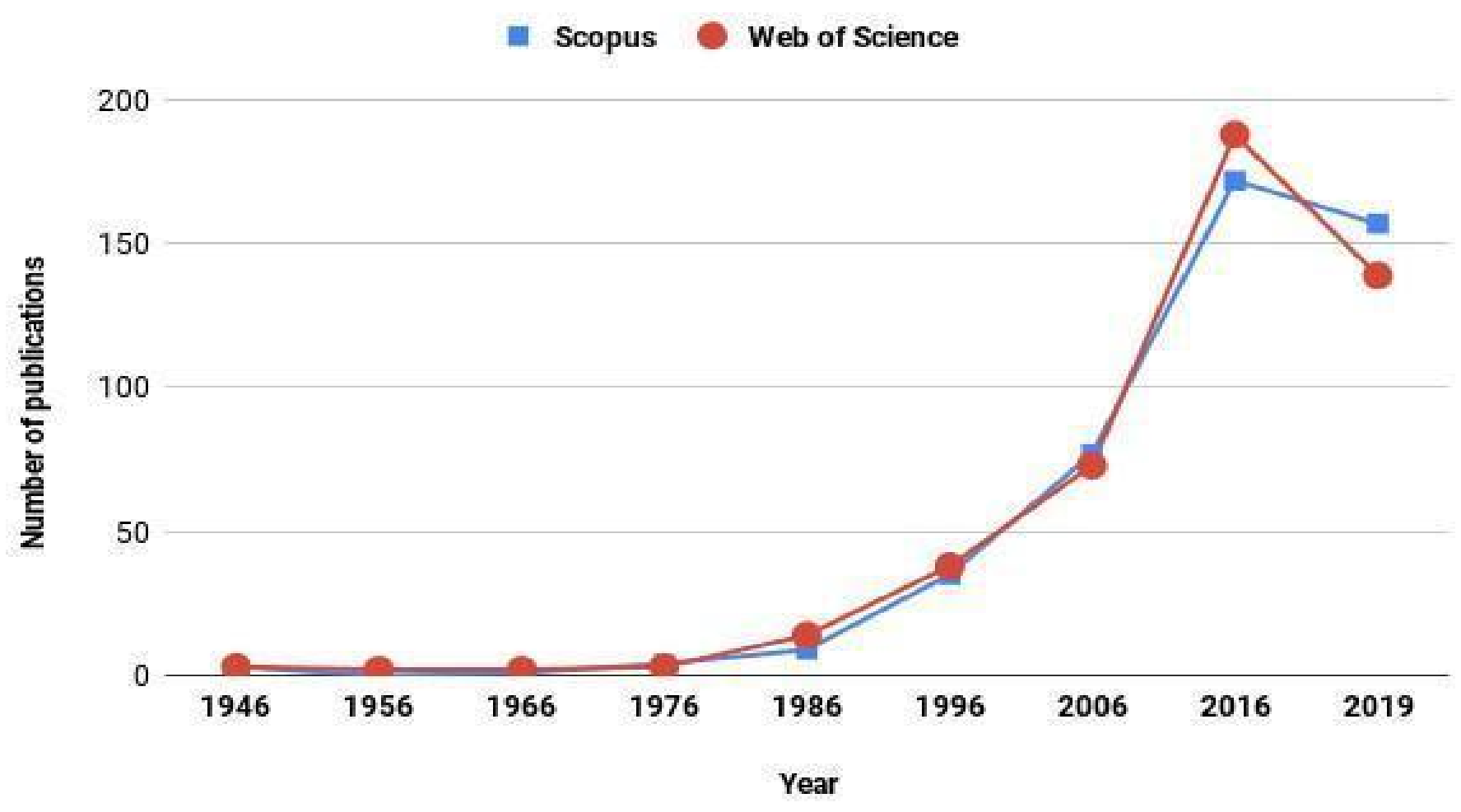

Fig. 1. Number of publications by year.

In general, the graph shows a significant growth in the number of publications over the years. This observation allows one to affirm that there is a progressive attention to the effects of NIR and to aspects relevant to the exposure and protection of the population, such as: development of technology for measurement of NIR; multidisciplinary studies on biological effects, leading to the creation and evolution of technical standards; proliferation of personal wireless communications systems, increasing apprehension and awareness motivating further studies to be carried out.

*C. Investment regions*

In order to analyze where the highest amount of studies in NIR is performed, the number of publications by country was surveyed to assess which nations have more significant research investment in this area. The results are shown in Fig. 2.

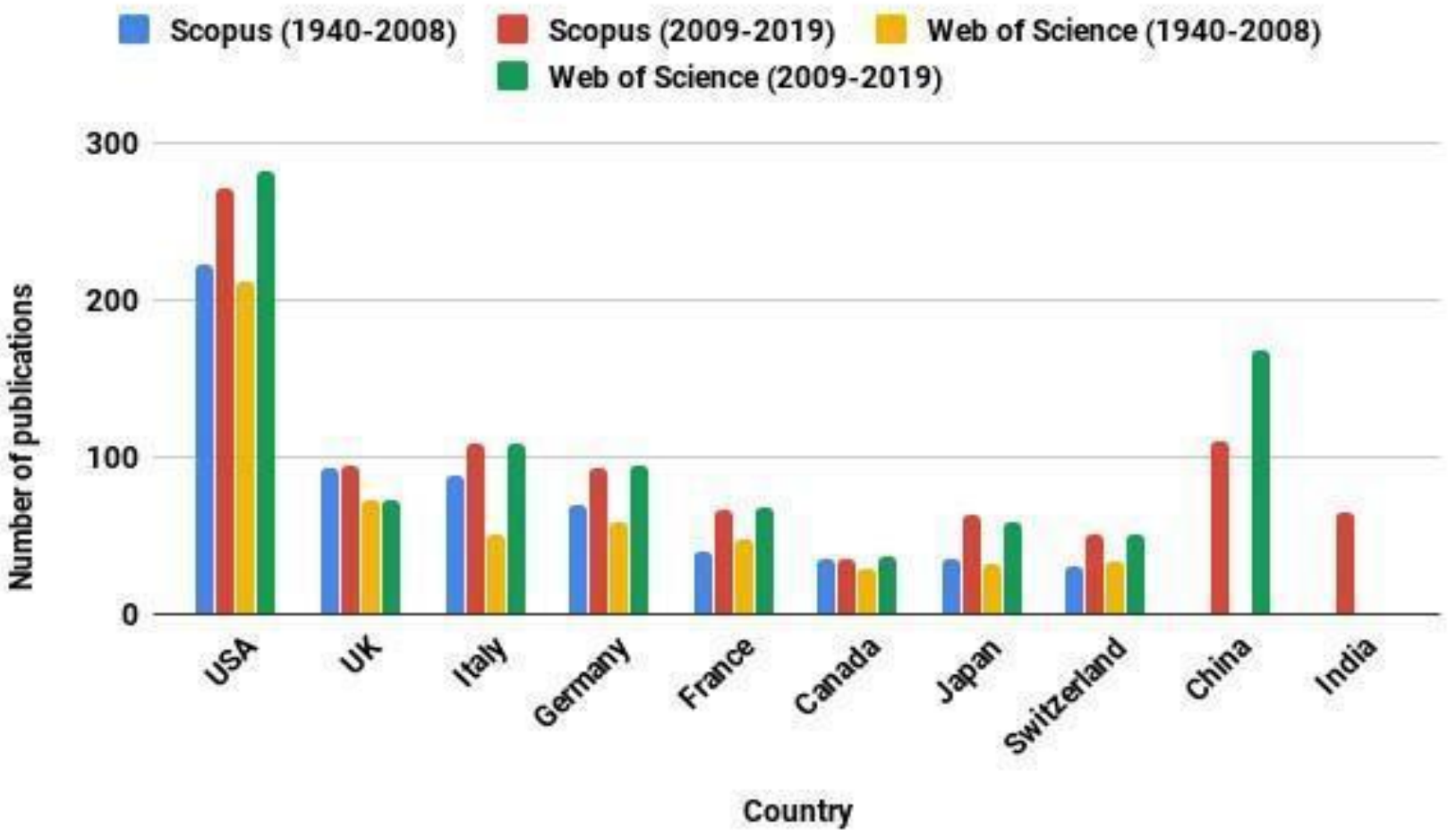

Fig. 2. Number of publications in different regions.

It is possible to observe that the United States lead the scientific production in the area, with significant growth from China and India in the last 10 years, especially as pointed out by the search made with the Scopus database. The other countries covered have maintained their numbers practically constant over the years, as pointed out by the two databases.

*D. Areas of Knowledge*

From the general perspective of growth in the number of publications, the next step of the bibliometric study was to identify in which areas of knowledge there is a higher relevance within NIR studies. Fig. 3 shows the number of publications by area, in the last ten years, considering the categories of WOS. The chart shows a predominance of the areas of engineering, radiology, physics, biophysics, public health and instrumentation, which is justified by the considerable variety of topics in which the study of NIR is relevant.

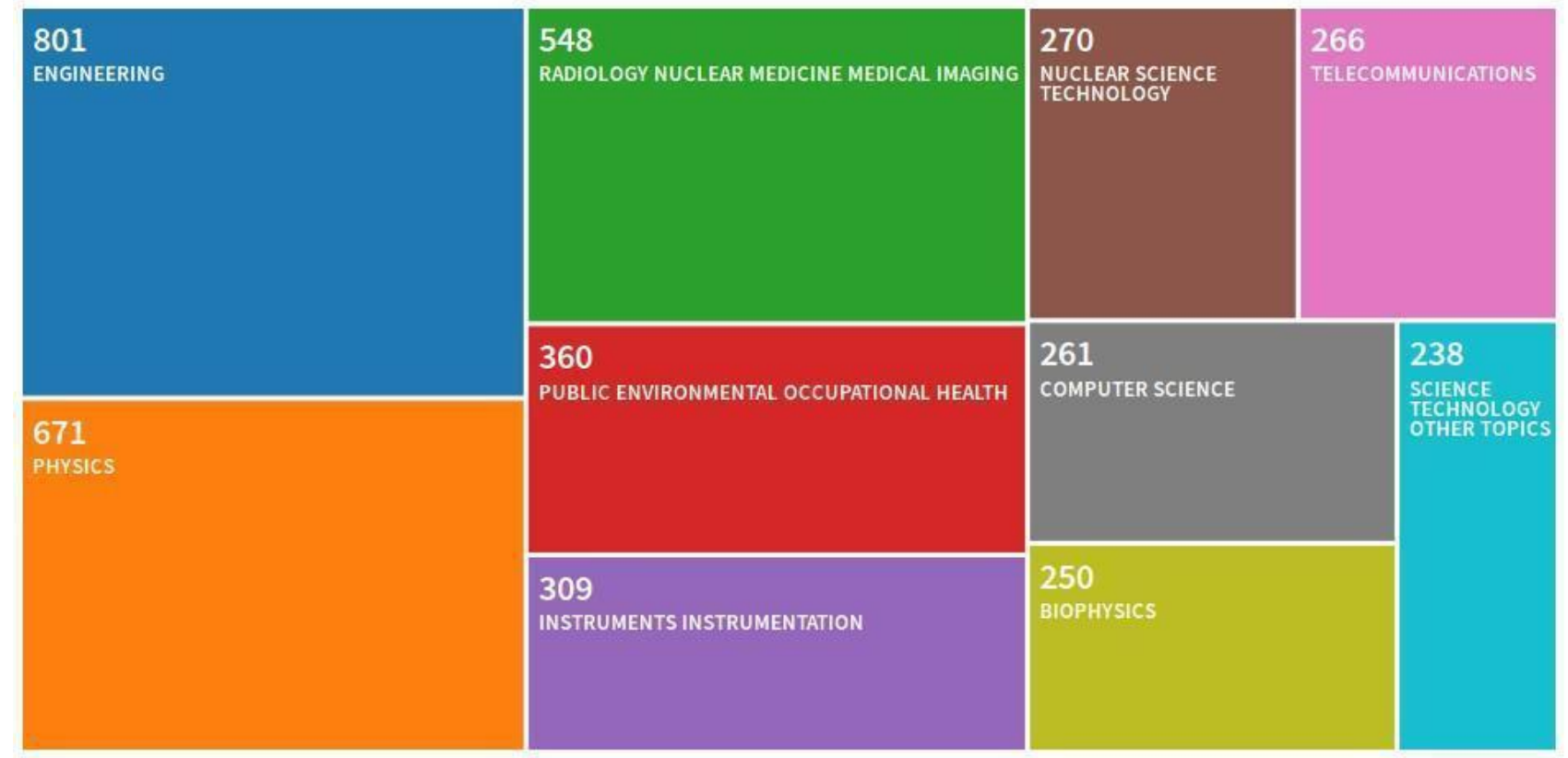

Fig. 3. Number of publications by area of knowledge.

To specify the relative trends in terms of application, a new search was conducted, in the Scopus database, for areas having articles with a relevant number of citations. For this, the 50 most cited articles in the last 10 and 3 years were considered. The 50 articles were classified in Table II according to the area they cover.

TABLE II. AREAS OF KNOWLEDGE WITH THE HIGHEST NUMBER OF PUBLICATIONS AMONG THE 50 MOST CITED.

| Area of knowledge | Publications (2009-2019) | Publications (2016-2019) |
|---|---|---|
| Radiology | 23 | 21 |
| Oncology | 8 | 10 |
| Astrophysics | 8 | 1 |
| Power Electronics | 5 | 4 |
| Electromagnetism | 4 | 4 |
| Sensoring | 1 | 2 |
| Biology | 1 | 2 |
| Chemistry | 0 | 2 |
| Physics | 0 | 2 |

In both periods, applications in the medical field, especially in radiology and oncology, have the highest number of publications. This may be reflected in a trend towards more investments and/or the increased fear of diseases linked to radiation. More recently, the areas of physics and chemistry have gained more relevance, showing the multidisciplinarity of the topic in question.

To confirm the trends related to the application areas, a search was performed for the keywords most used in the mentioned periods, as shown in Table III.

TABLE III. MOST FREQUENT KEYWORDS.

| **Keywords** | **Publications (2009-2019)** | **Publications (2016-2019)** |
|---|---|---|
| Human | 810 | 278 |
| Electromagnetic Fields | 349 | 131 |
| Radiation Protection | 221 | 99 |
| Radiation Exposure | 211 | 71 |
| Radiation Shielding | 139 | 49 |
| Radio Waves | 129 | 37 |
| Electromagnetic Field Effects | 126 | 52 |
| Environmental Exposure | 126 | 29 |

By analyzing Table III, it is possible to notice a correspondence between the results of both periods, i.e., the same sequence of most frequent keywords. We also observe the affinity in the topics with predominance when comparing Tables III and II, pointing that the main concern is concentrated on the conservation of the human quality of life.

## III. EXPOSURE LIMITS

Given the indication of the influence of NIR for the development of diseases and other injuries in the human body [2], [19]–[22], although studies are not conclusive, as well as the amount of publications in the medical field about the subject, several multidisciplinary organizations in the world have established recommendations regarding the limits of human exposure to this radiation. In the World Health Organization, the body responsible for this task at the international level, as previously mentioned, is ICNIRP [1]. This entity determined the limits of exposure to EMF and power density of the equivalent plane wave for two categories: (i) occupational populations (individuals who work near NIR sources and are aware of it); and (ii) general populations (individuals who do not work near NIR sources and are unaware of the sources and their possible effects).

In Brazil, ANATEL [4], [3] is the institution responsible for establishing resolutions regarding the limit of human exposure to NIR, based on the data and studies released by ICNIRP. Tables IV, V, VI and VII present the occupational and general exposure limits of ICNIRP and ANATEL. In the tables, $f$ denotes the wave frequency in Hertz (Hz), $E$ the electric field strength in V/m, $H$ the magnetic field strength in A/m, $B$ the density of magnetic flux in microteslas and $S$ the power density of the equivalent plane wave in W/m$^2$. It is important to note that the unit of $f$ for the calculation of $E$, $H$, $B$ and $S$ must follow the multiple of Hz indicated in the frequency range under analysis.

TABLE IV. EXPOSURE LIMITS FOR THE OCCUPATIONAL POPULATION (ICNIRP) [1].

| Frequency range | *E* field | *H* field | *B* field | *S* field |
|---|---|---|---|---|
| Up to 1Hz | - | $1.63\times10^5$ | $2\times10^5$ | - |
| 1Hz to 8Hz | 20000 | $1.63\times10^5/f$ | $2\times10^5/f$ | - |
| 8Hz to 25Hz | 20000 | $2\times10^4/f$ | $2.5\times10^4$ | - |
| 0.025 to 0.82kHz | $500/f$ | $20/f$ | $325/f$ | - |
| 0.82 to 65kHz | 610 | 24.4 | 30.7 | - |
| 0.065 to 1MHz | 610 | $1.6/f$ | $2/f$ | - |
| 1 to 10MHz | $610/f$ | $1.6/f$ | $2/f$ | - |
| 10 to 400MHz | 61 | 0.16 | 0.2 | 10 |
| 400 to 2000MHz | $3\sqrt{f}$ | $0.008\sqrt{f}$ | $0.01\sqrt{f}$ | $f/40$ |
| 2 to 300GHz | 137 | 0.36 | 0.45 | 50 |

TABLE V. EXPOSURE LIMITS FOR THE GENERAL POPULATION (ICNIRP) [1].

| Frequency range | *E* field | *H* field | *B* field | *S* field |
|---|---|---|---|---|
| Up to 1 Hz | - | $3.2\times10^4$ | $4\times10^4$ | - |
| 1 to 8 Hz | 10000 | $3.2\times10^4/f^2$ | $4\times10^4/f$ | - |
| 8 to 25 Hz | 10000 | $4000/f$ | $5000/f$ | - |
| 0.025 to 0.82 kHz | $250/f$ | $4/f$ | $5/f$ | - |
| 0.82 to 3 kHz | $250/f$ | 5 | 6.25 | - |
| 3 to 150 kHz | 87 | 5 | 6.25 | - |
| 0.15 to 1 MHz | 87 | $0.73/f$ | $0.92/f$ | - |
| 1 to 10 MHz | $87\sqrt{f}$ | $0.73/f$ | $0.92/f$ | - |
| 10 to 400 MHz | 28 | 0.073 | 0.092 | - |
| 400 to 2000 MHz | $1375\sqrt{f}$ | $0.0037\sqrt{f}$ | $0.0046\sqrt{f}$ | $f/2000$ |
| 2 to 300GHz | 61 | 0.16 | 0.2 | 10 |

TABLE VI. EXPOSURE LIMITS FOR THE OCCUPATIONAL POPULATION (ANATEL) [4].

| Frequency range | *E* field | *H* field | *S* field |
|---|---|---|---|
| 8.3 to 65kHz | 170 | 24.4 | - |
| 0.065 to 3.6MHz | 170 | $1.6/f$ | - |
| 3.6 to 10MHz | $610/f$ | $1.6/f$ | - |
| 10 to 400MHz | 61 | 0.16 | 10 |
| 400 to 2000MHz | $3\sqrt{f}$ | $0.008\sqrt{f}$ | $f/40$ |
| 2 to 300GHz | 137 | 0.36 | 50 |

TABLE VII. EXPOSURE LIMITS FOR THE GENERAL POPULATION (ANATEL) [4].

| Frequency range | *E* field | *H* field | *S* field |
|---|---|---|---|
| 8.3 to 150kHz | 83 | 5 | - |
| 0.15 to 1MHz | 83 | $0.73/f$ | - |
| 1 to 10MHz | $87\sqrt{f}$ | $0.73/f$ | - |
| 10 to 400MHz | 28 | 0.073 | 2 |
| 400 to 2000MHz | $1.375\sqrt{f}$ | $0.037\sqrt{f}$ | $f/200$ |
| 2 to 300GHz | 137 | 0.16 | 10 |

As shown in Tables IV, V, VI and VII, there is a significant difference between the limits at lower frequencies, in addition to the fact that ANATEL does not present the limits for field *B*.

*A. Brazilian NIR Measurement Recommendations*

ANATEL has a series of practical procedures for measuring the parameters necessary to assess human exposure to NIR. The main metrics are fields *E* and *H*, and the study is normally done in the Fraunhofer region of the stations under analysis [4].

Probe equipment with quasi-isotropic characteristics and polarization switching is more suitable to this task, although antennas with a more directive radiation diagram can also be used. It is essential that the used apparatus covers the entire frequency range of the waves that may be being emitted in the observed domain.

Regarding the recommended measurement spots for the analysis of the general population exposure, the following situations are of interest [4]:

- At a point in the place of maximum exposure resulting from the transmitting station;
- An area of great circulation of people;
- At three points in the direction of maximum radiation from the emitting systems from the base of the antenna support structure (one point at a distance less than 50m, another at a distance between 50m and 150m and another over 150m); and – at one point in the Assessment Domain Boundary (ADB) [23].

The ADB, also called the research domain, when it comes to omnidirectional transmitting antennas, has the shape of a cylinder of D radius and a height of 3.5 + $H_b$, calculated by expressions (1) and (2) (for frequencies above 30 MHz) [4]:

$$D = 1.3\sqrt{\sum_i \frac{EIRP_i}{S_{lim,i}}} \quad . \tag{1}$$

$$H_b = max(3.5; tg(\alpha)). \tag{2}$$

where $EIRP_i$ denotes the effective isotropically radiated power for a frequency *i*, $S_{lim,i}$ the limit power density for a frequency *i*, and $\alpha$ the highest tilt among all the transmitting antennas of the support structure. When the emission is less than 30 MHz, the ADB is a circle of radius *r*, defined by Tables VIII and IX.

TABLE VIII. DEFINITION OF ADB FOR FREQUENCIES BELOW 30 MHZ (OCCUPATIONAL POPULATION) [4].

| Frequency range | *r* (in meters) |
|---|---|
| 0.525 to 3.6 MHz | $0.076\sqrt{f \times EIRP}$ |
| 3.6 to 10 MHz | $0.04f\sqrt{EIRP}$ |
| 10 to 30 MHz | $0.404\sqrt{EIRP}$ |

TABLE IX. DEFINITION OF ADB FOR FREQUENCIES BELOW 30 MHZ (GENERAL POPULATION) [4].

| Frequency range | *r* (in meters) |
|---|---|
| 0.525 to 3.6 MHz | $0.162\sqrt{f \times EIRP}$ |
| 3.6 to 10 MHz | $0.04\sqrt{f^3 EIRP}$ |
| 10 to 30 MHz | $0.882\sqrt{EIRP}$ |

For the occupational population, the recommended spots for measurement are:

- At a point at the place of maximum radiation exposure;
- At a point where there is a greater flow of workers.

When the maximum level of the electric field strength exceeds 50% of the limit for the frequency under analysis, attendance should be demonstrated based on a time average (using the measurement time of 6 minutes, for example) of at least 3 vertical measurements, as illustrated in Fig. 4. The spatial mean is calculated by

$$E_{spatial_mean} = \sqrt{\frac{E_1^2 + E_2^2 + E_3^2}{3}} \quad . \tag{3}$$

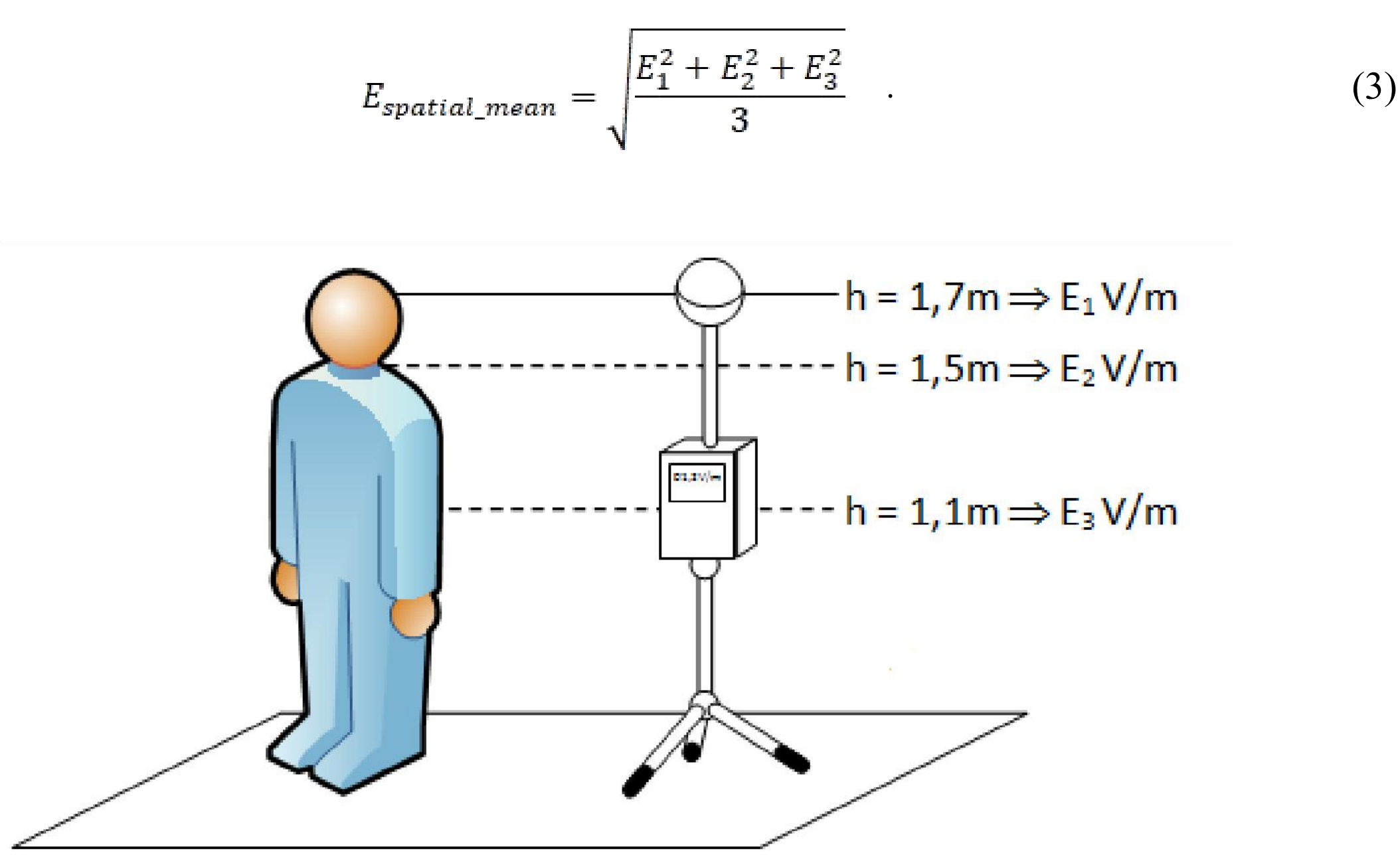

Fig. 4. Illustration of measurements for spatial average determination (adapted from [4]).

Summarizing the Brazilian efforts on NIR legislation, in 2002, ANATEL defined in [24] the NIR limits and regulations; In 2018, the rules were replaced by [3], with some amendment in 2019 by [4], which was previously described. These updates on the Brazilian legislation regarding NIR levels have motivated the measurement study (presented in the next section), since many telecommunication services are available for customer use in areas surrounding a number of buildings, including 4G and pre-5G communications, digital TV and some Wi-Fi versions that operate in 5 GHz frequency.

## IV. MEASUREMENTS IN NORTHEAST BRAZIL

This section reports an investigation of NIR levels by presenting the results of measurement rounds conducted in four different locations during a period of 24 hours in each spot. Recorded electric field intensities will be then compared to international and local Brazilian regulations (since they establish the same exposure limit for the analyzed frequencies), as well as to theoretical field intensity calculations, finally stating whether NIR exposure at the considered environments is below or above the limits established by these regulations.

### *A. Measurement Setup*

Regarding the utilized apparatus, all measurement rounds were conducted with a calibrated Wavecontrol MonitEM equipment borrowed from ANATEL, with frequencies ranging from 100 kHz to 8 GHz with quasi-isotropic radiation diagram and capability of three-axis measurements. Thus, it includes an important range of legacy and modern communications services. The probe, however, does not allow one to scrutinize the contribution of field intensity for each analyzed frequency. The recorded levels were stored with a sampling frequency of one sample per second. Fig. 5 shows a measured setup in a real acquisition round.

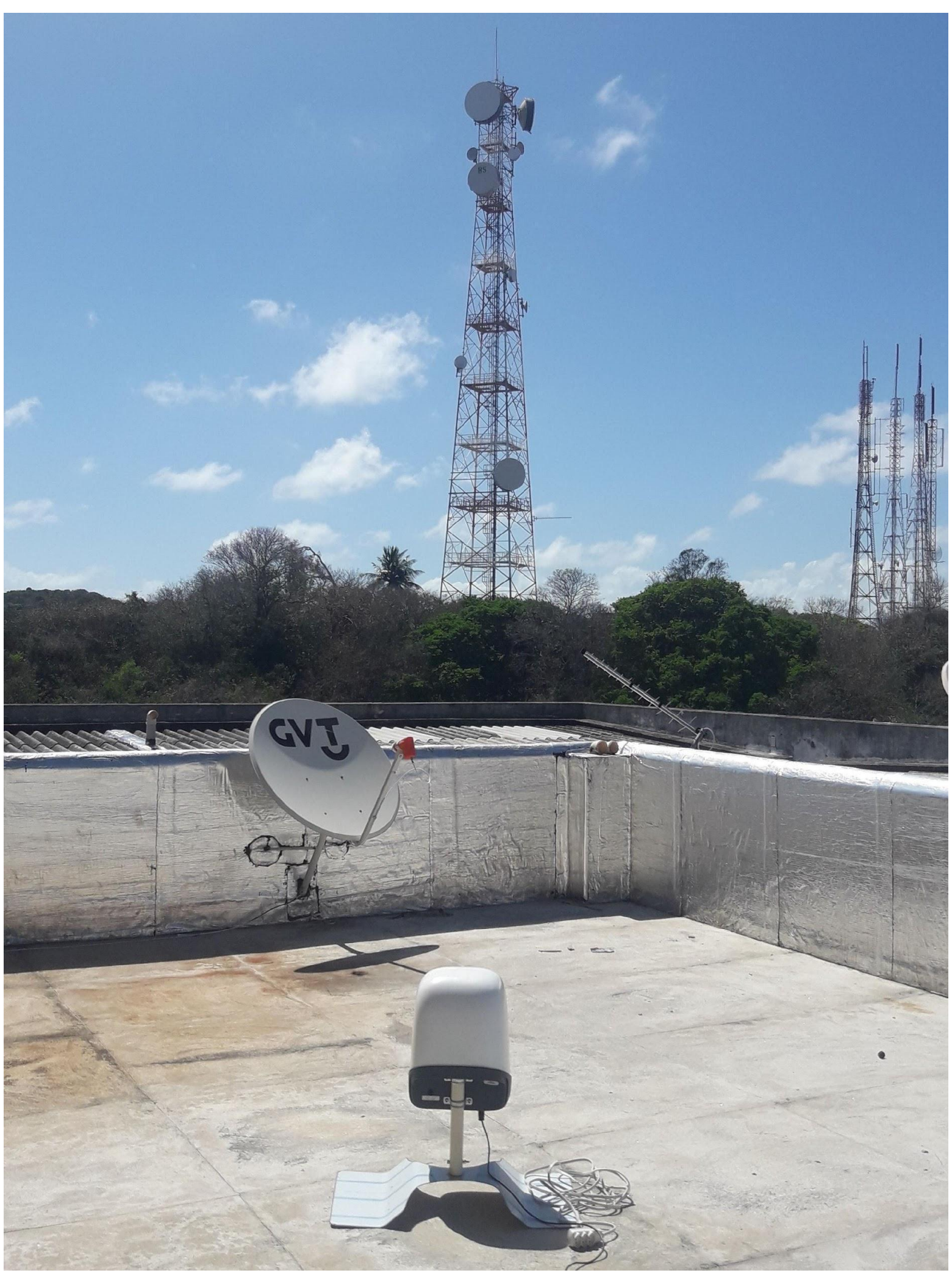

Fig. 5. Example of measurement setup and site.

*B. Measurement Points*

The data acquisition was performed in 4 different locations in the proximity of an antenna spot at skyline level, as presented in Fig. 6. We claim it could represent a source of high NIR levels for a considerable amount of people who live and work around the area, especially because they are close to several broadcast and cellular sites. Extending the information of Fig. 6, the geographical coordinates of buildings where the measurements were conducted are identified as (with distances from Google Maps [25]):

- P1: Alto do Tirol Apartment House (5◦47'53.2"S, 35◦11'49.8"W), 197 m from the antenna spot;
- P2: Portal do Tirol Apartment House (5◦47'51.1"S, 35◦11'50.5"W), 153 m from the antenna spot;
- P3: Casablanca Condominium (5◦47'39.4"S, 35◦11'52.2"W), 290 m from the antenna spot; and
- P4: Athletical Association of the Bank of Brazil (AABB) (5◦47'47.5"S, 35◦11'55.9"W), 265m from the antenna spot.

Respectively, the measurement rounds were held on September 12th (beginning at 15:43:34), October 14th (beginning at 09:18:46), November 5th (beginning at 08:55:22) and November 29th (beginning at 10:52:47). In Fig. 6, the measurement locations are marked with blue spots, while one of the closest antenna is marked with a green one.

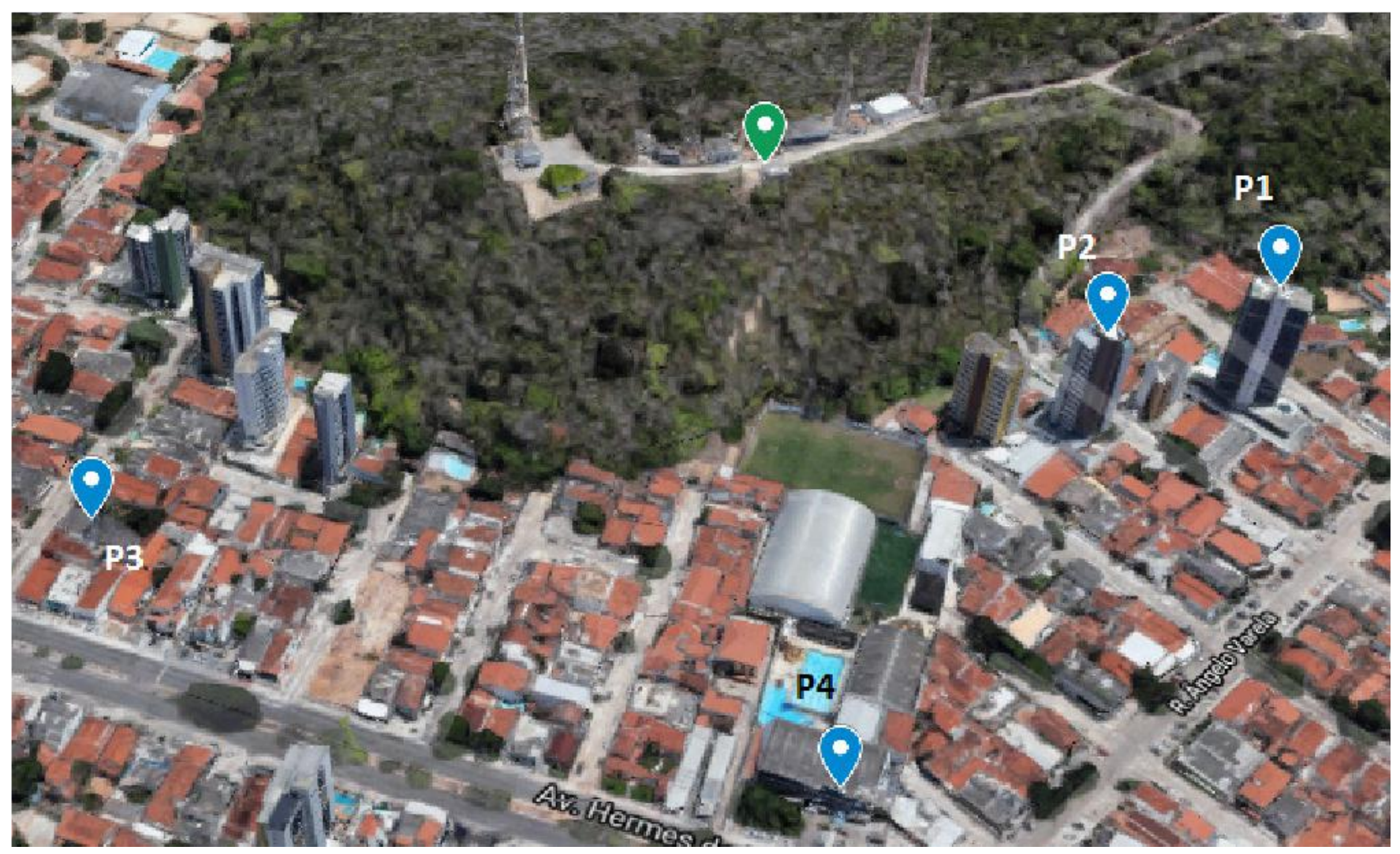

Fig. 6. Measurement locations, identifying the closest antenna and its covered area. Source: Google Maps [25].

*C. Result Presentations and Discussions*

Information about frequency assignment and localization of service antennas is publicly provided by ANATEL. According to the available frequency plan, the analyzed antenna spot is used for digital TV broadcasting and the lowest utilized frequency is 470 MHz. Thus, following Table VII, the strictest exposure limit is 29.81 V/m ( $1.375\sqrt{f}$ $1.375\sqrt{f}$ ). Based on this value and the conducted measurements in each round, the results of Table XI were achieved, where $\mu$ denotes the mean value, and $\sigma$ the standard deviation for each measurement round over time.

TABLE XI. MEASUREMENT RESULTS IN V/M.

| Measurement point | $\mu$ | $\sigma$ | Min. Value | Max. Value | % of Limit[1] |
|---|---|---|---|---|---|
| Portal do Tirol (P2) | 12.88 | 0.76 | 11.11 | 15.03 | 50.42 |
| Alto do Tirol (P1) | 9.21 | 0.15 | 8.70 | 9.51 | 31.90 |
| AABB (P4) | 5.19 | 0.23 | 2.75 | 5.79 | 19.42 |
| Casablanca (P3) | 4.39 | 0.11 | 3.83 | 4.65 | 15.60 |

Fig. 7 shows Electric field intensity measurements through time. We can observe that the Portal do Tirol building reached the strongest electric field intensity values, thus, the highest percentage of the national exposure limit. This qualitative result is expected, given that it is the spot located the closest to the transmitting antenna. Such measurement point also presented the highest standard deviation.

For the other buildings, the behavior of field intensity is practically constant for the whole measuring time, with considerably smaller variations compared to Portal do Tirol. Their mean, maximum and minimum values can be explained by the distance from the measurement spots to the antenna towers.

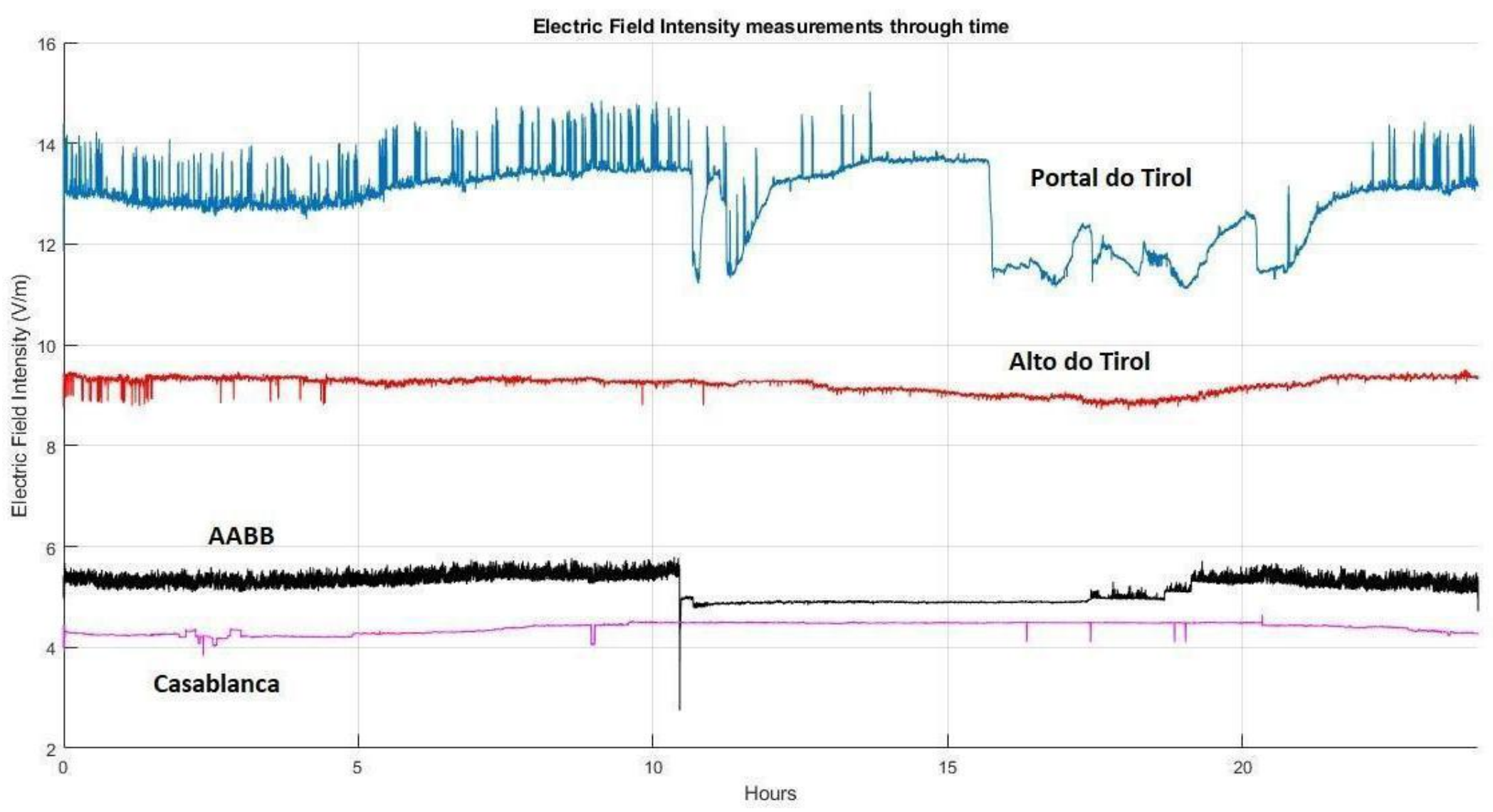

Fig. 7 Electric field intensity measurements through time.

D. *Theoretical Discussions*

In the given scenario, it is of interest to investigate the contribution of a number of antennas, each one operating at a different frequency, in the electric field intensity at the measurement points.

[1] Although our measurements are wideband, the limit corresponds to 470 MHz (the worst case).

International guidelines [1] and current Brazilian regulations [4] consider propagation in free space. From equation (B.6) in [4]:

$$|E|^2 = S \times 377. \quad (4)$$

Here, $|E|$ denotes the absolute value of the electrical field intensity (in V/m), S is the power density (absolute value of the Poynting vector in the radial direction, in $W/m^2$) and 377 being an approximation of $120\pi$, accounting for the characteristic impedance of free space. According to equation (B.5) in [4]:

$$S = \frac{EIRP}{4\pi R^2} \times F(\theta, \varphi) = \frac{P_t G_t}{4\pi r^2} \times F(\theta, \varphi), \quad (5)$$

where $EIRP$ means the effective isotropic radiated power from an antenna, $R$ is the distance of the antenna center of phase to the point of interest (in meters), $F(\theta, \varphi)$ the maximum gain reduction factor as a function of radiation direction, $P_t$ the transmitting power at the antenna output terminals (in watts) and $G_t$ the gain of the transmitting antenna in a given direction (in linear scale, referred to the isotropic reference). From the previous equations, it is then possible to achieve:

$$|E| = \frac{\sqrt{30 P_t G_t \times F(\theta, \varphi)}}{R}. \quad (6)$$

The following analysis considers conservative estimations. Thus, in the context of NIR level estimations, the most critical scenario will be taken, meaning that, in the next calculations, $F(\theta, \varphi)$ will be taken as equal to 1.

Data about the antennas located at the considered transmitting spot are made public by ANATEL. These include information such as operation frequency and power, transmission line length and losses, gain and polarization.

Based on this, it is possible to theoretically determine the contribution of the antennas yielding the strictest limitation on electric field intensity according to the current regulations, giving a conservative estimate. Their operation parameters are available in Table XII.

TABLE XII: ANTENNA OPERATION PARAMETERS.

| **Frequency (MHz)** | **Transmitting power (W)** | **Transmission line length (m)** | **Attenuation (dB/100m)** | **Other losses (dB)** | **Maximum gain (dBd)** | **Output transmitting power (W)** |
|---|---|---|---|---|---|---|
| 470 | 2400 | 50 | 1.48 | 0.5 | 11.52 | 1803.85 |
| 515 | 2200 | 70 | 1.63 | 0.5 | 8.35 | 1507.65 |
| 533 | 2200 | 70 | 1.58 | 0.5 | 10.6 | 1519.85 |
| 569 | 3400 | 70 | 2.51 | 0.5 | 11.98 | 2022.09 |

The final theoretical value of the electric field intensity is given by:

$$|E_{final}| = \sqrt{|E_{470}|^2 + |E_{515}|^2 + |E_{533}|^2 + |E_{569}|^2}. \quad (7)$$

By calculating $|E_{final}|$ at each point of interest, their values are shown in Table XIII.

TABLE XIII: ESTIMATED ELECTRIC FIELD INTENSITIES.

| Spot | $\|E_{final}\|$ (V/m) |
|---|---|
| P1 | 10.38 |
| P2 | 13.73 |
| P3 | 7.06 |
| P4 | 7.72 |

It can be seen that these contributions are higher than the average value of the measured electric field intensities at each point over the time period in question. When the measurement campaigns took place, there were 16 antennas located at the analyzed spot. This performed estimation, however, considered the most restrictive scenario considering NIR levels, according to current Brazilian regulations [4], and have generated a more intense electric field in comparison to the measured values. Thus, these calculations reflect a more rigorous setting than a more extensive and exhaustive analysis would suggest.

The difference between the calculated and measured values can be addressed by a variety of factors, including but not limited to: the use of the simplistic free space propagation model in an urban scenery, which does not take into account propagation mechanisms causing scattering and attenuation of the propagating wave, like multiple diffractions and reflections; possible imprecision in the database of radiation system elements of the considered antennas; the use of a very conservative gain value (the maximum one) for all antennas, in the absence of their radiation pattern.

## V. FINAL REMARKS AND FURTHER INVESTIGATIONS

The results of the prospective analysis carried out in this work show a considerable increase in the number of publications on NIR over the years in several areas of knowledge, from topics in electromagnetism to more specific areas of medicine. The temporal analysis revealed the countries that have attained greater relevance in the number of publications in the world scientific scenario, such as China and India, especially in recent years. In addition, the growing number of productions in recent decades attests to a concern with the current spectrum usage, given the notable expansion and rapid diversification in associated services. In addition, it is evident that NIR discussions also appear in other contexts since this radiation is used for some purposes, such as structures for energy transfer in power systems [26], radiation processing in plants [27], and development of drivers for the hadron collider [28].

Based on several discussions about the effects of radiation in humans, international guidelines of NIR exposure were established by ICNIRP, serving as a reference for some country institutions, such as ANATEL, in the case of Brazil. The latter also describes a series of procedures for assessing the exposure of occupational and general populations to radiation in several cases, in terms of the frequency range under analysis, and configuration of the transmitting antennas.

The measurement rounds conducted in four different buildings in Natal, Brazil, serve as an example for assessment of general population exposure to NIR. Although none of them have surpassed the recommended exposure levels, it is important to note that they can rise in a not so distant future, since the number of telecommunication services tend to grow rapidly. Furthermore, it is possible to speculate that electric field intensities, such as the one reached at Portal do Tirol, could be higher in the following years.

The achieved results allow a broader view for the perspective of developing studies in the coming years. As a continuation of this work, it is possible to include research on situations subjected to important NIR levels and concentration of people, like indoor environments using small cell densification, as well as the study of the intensity of NIR in sources that operate in the frequency range of millimeter waves, for example.

Another suggestion is the conduction of measurement rounds in bigger urban centers, where exposure levels might be much stronger than the ones reached in Natal, Brazil.

ACKNOWLEDGEMENT

This study was financed in part by the Coordenação de Aperfeiçoamento de Pessoal de Nível Superior Brasil (CAPES) - Finance Code 001. Authors would like to express their gratitude to ANATEL for providing all equipment for this research, as well as the valuable directions about the measurement procedures.